\documentclass[11pt]{article}%
\usepackage{amsmath}
\usepackage{graphicx}%
\usepackage{amsfonts}%
\usepackage{amssymb}
\newtheorem{theorem}{Theorem}

\newtheorem{example}[theorem]{Example}

\newtheorem{proposition}[theorem]{Proposition}
\newtheorem{remark}[theorem]{Remark}

\newenvironment{proof}[1][Proof]{\textbf{#1.} }{\ \rule{0.5em}{0.5em}}

\begin{document}

\title{Quaternionic factorization of the Schroedinger operator and\ its applications
to some first order systems of mathematical physics}
\author{Viktor G. Kravchenko*, Vladislav V. Kravchenko**\\\ *Faculdade de Ciencias y Tecnologia\\\ \ Universidade do Algarve,\ Campus de Gambelas\\\ \ 8000 Faro, \ \ PORTUGAL\\\ \ e-mail: vkravch@ualg.pt\\\ **Depto. de Telecomunicaciones \\\ \ \ SEPI ESIME Zacatenco \\\ \ \ Instituto Polit\'{e}cnico Nacional\\\ \ \ Av. IPN\ S/N,\ Edif. 1\\\ \ \ C.P. 07738, D.F.,\ \ MEXICO\\\ \ \ vkravchenko@ipn.mx}
\maketitle

\section{Introduction}

We consider the following first order systems of mathematical physics.

\begin{enumerate}
\item The Dirac equation with scalar potential.

\item The Dirac equation with electric potential.

\item The Dirac equation with pseudoscalar potential.

\item The system describing non-linear force free magnetic fields or Beltrami
fields with nonconstant proportionality factor.

\item The Maxwell equations for slowly changing media.

\item The static Maxwell system.
\end{enumerate}

We show that all this variety of first order systems reduces to the equation
\begin{equation}
Df+f\cdot\overrightarrow{\alpha}=0, \label{1}%
\end{equation}
where $D$ is the Moisil-Theodoresco operator (introduced by Hamilton) acting
on biquaternion valued functions $f$ according to the rule $Df=\sum_{k=1}%
^{3}e_{k}\partial_{k}f$, $\partial_{k}=\frac{\partial}{\partial x_{k}}$,
$e_{k}$ are standard quaternionic imaginary units; the function $f$ of real
variables $x_{1}$, $x_{2}$, $x_{3}$ has the form $f=\sum_{k=0}^{3}f_{k}e_{k}$,
$f_{k}\in C$, $k=0,1,2,3$ and $\overrightarrow{\alpha}$ is a purely vectorial
biquaternion valued function. We reduce the solution of equation (\ref{1}) to
the solution of a Schr\"{o}dinger equation with biquaternionic potential. In
some important situations the biquaternionic potential can be diagonalized and
converted into scalar potentials.

\section{Notations}

We will consider the algebra $\mathbb{H}(\mathbb{C})$ of complex quaternions
or biquaternions which have the form $q=\sum_{k=0}^{3}q_{k}e_{k}$ where
$\{q_{k}\}\subset\mathbb{C}$, $e_{0}$ is the unit and $\{e_{k}|\quad
k=1,2,3\}$ are the quaternionic imaginary units, that is the standard basis
elements possessing the following properties:%

\[
e_{0}^{2}=e_{0}=-e_{k}^{2};\;e_{0}e_{k}=e_{k}e_{0}=e_{k},\quad k=1,2,3;
\]%

\[
e_{1}e_{2}=-e_{2}e_{1}=e_{3};\;e_{2}e_{3}=-e_{3}e_{2}=e_{1};\;e_{3}%
e_{1}=-e_{1}e_{3}=e_{2}.
\]
We denote the imaginary unit in $\mathbb{C}$ by $i$ as usual. By
definition\ \ $i$\ commutes with \ $e_{k}$, $k=0,1,2,3$.

The vectorial representation of a complex quaternion will be used. Namely,
each complex quaternion $q$ is a sum of a scalar $q_{0}$ and of a vector
$\overrightarrow{q}$:
\[
q=\operatorname*{Sc}(q)+\operatorname*{Vec}(q)=q_{0}+\overrightarrow{q},
\]
where $\overrightarrow{q}=\sum_{k=1}^{3}q_{k}e_{k}$. The purely vectorial
complex quaternions ($\operatorname*{Sc}(q)=0$) are identified with vectors
from $\mathbb{C}^{3}$. The following conjugation operations will be needed.
The quaternionic conjugation is defined as follows $\overline{q}%
=q_{0}-\overrightarrow{q}$, the complex conjugation: $q^{\ast}%
=\operatorname{Re}q-i\operatorname{Im}q$ and the following involutive
operation $q^{(k)}=e_{k}q\overline{e_{k}}$ which changes signs of two
components, for example, $q^{(1)}=-e_{1}qe_{1}=q_{0}e_{0}+q_{1}e_{1}%
-q_{2}e_{2}-q_{3}e_{3}$.

By $M^{p}$ we denote the operator of multiplication by a complex quaternion
$p$ from the right-hand side: $M^{p}q=q\cdot p$.

We will intensively use the fact that the algebra of complex quaternions
contains a subset of zero divisors $\mathfrak{S}$ which are characterized by
the equality $q_{0}^{2}=\overrightarrow{q}^{2}$, where $\overrightarrow{q}%
^{2}=-\left\langle \overrightarrow{q},\overrightarrow{q}\right\rangle $, or
equivalently $q^{2}=2q_{0}q$. Hence if $q\in\mathfrak{S}$ and $q_{0}=1/2$ then
$q$ is an idempotent. Using this fact we introduce the following
multiplication operators $P_{k}^{\pm}=\frac{1}{2}M^{(1\pm ie_{k})}$,
$k=1,2,3$. It is easy to see that for each $k$ the operators $P_{k}^{+}\ $and
$P_{k}^{-}$ represent a pair of mutually complementary, orthogonal projection
operators on the set of $\mathbb{H}(\mathbb{C})$-valued functions. More
information on the structure of the algebra of complex quaternions can be
found for example in \cite{AQA} or \cite{KSbook}.

Let $f$ be a complex quaternion valued differentiable function of
$\mathbf{x}=(x_{1},x_{2},x_{3})$. Denote
\[
Df=\sum_{k=1}^{3}e_{k}\frac{\partial}{\partial x_{k}}f.
\]
This expression can be rewritten in vector form as follows%

\[
Df=-\operatorname*{div}\overrightarrow{f}+\operatorname*{grad}f_{0}%
+\operatorname*{rot}\overrightarrow{f}.
\]
That is, $\operatorname*{Sc}(Df)=-\operatorname*{div}\overrightarrow{f}$ and
$\operatorname*{Vec}(Df)=\operatorname*{grad}f_{0}+\operatorname*{rot}%
\overrightarrow{f}$. Let us notice that $D^{2}=-\Delta$. The operator
$D+M^{\overrightarrow{\alpha}}$ we will denote also by $D_{\overrightarrow
{\alpha}}$.

Let us introduce an auxiliary notation $\widetilde{f}:=f(x_{1},x_{2},-x_{3})$.
The domain $\widetilde{\Omega}$ is assumed to be obtained from the domain
$\Omega\subset\mathbb{R}^{3}$ by the reflection $x_{3}\rightarrow-x_{3}$.

\section{First order systems reducing to equation (\ref{1})}

\subsection{The Dirac equation with scalar potential}

Let $\mathcal{D}$ denote the classic Dirac operator for a free particle with a
specified energy $\omega\in\mathbb{R}$%
\[
\mathcal{D}=i\omega\gamma_{0}+\sum_{k=1}^{3}\gamma_{k}\partial_{k}+im.
\]
Here $\gamma_{j},$ $j=0,1,2,3$ are usual $\gamma$-matrices (see, e.g.,
\cite{BD}) and $m\in\mathbb{R}$.

The Dirac operator with scalar potential has the form (see, e.g.,
\cite{Thaller})%
\[
\mathcal{D}^{sc}=\mathcal{D}+i\varphi_{sc}I,
\]
where $\varphi_{sc}$ is a scalar real-valued function of $\mathbf{x}$ and $I$
denotes the identity operator.

In \cite{Krbag} (see also \cite{AQA}, \cite{KSbook}) the following
transformation was introduced. A function $\Phi:\Omega\subset\mathbb{R}%
^{3}\rightarrow\mathbb{C}^{4}$ is transformed into a function $F:\widetilde
{\Omega}\subset\mathbb{R}^{3}\rightarrow\mathbb{H(C)}$ by the rule
\[
F=\mathcal{A}[\Phi]:=\frac{1}{2}\left(  -(\widetilde{\Phi}_{1}-\widetilde
{\Phi}_{2})e_{0}+i(\widetilde{\Phi}_{0}-\widetilde{\Phi}_{3})e_{1}%
-(\widetilde{\Phi}_{0}+\widetilde{\Phi}_{3})e_{2}+i(\widetilde{\Phi}%
_{1}+\widetilde{\Phi}_{2})e_{3}\right)  .
\]
The inverse transformation $\mathcal{A}^{-1}$ is defined as follows
\[
\Phi=\mathcal{A}^{-1}[F]=(-i\widetilde{F}_{1}-\widetilde{F}_{2},-\widetilde
{F}_{0}-i\widetilde{F}_{3},\widetilde{F}_{0}-i\widetilde{F}_{3},i\widetilde
{F}_{1}-\widetilde{F}_{2}).
\]
Let us present the introduced transformations in a more explicit matrix form
which relates the components of a $\mathbb{C}^{4}$-valued function$\ \Phi$
with the components of an $\mathbb{H(C)}$-valued function $F$:
\[
F=\mathcal{A}[\Phi]=\frac{1}{2}\left(
\begin{array}
[c]{rrrr}%
0 & -1 & 1 & 0\\
i & 0 & 0 & -i\\
-1 & 0 & 0 & -1\\
0 & i & i & 0
\end{array}
\right)  \left(
\begin{array}
[c]{c}%
\widetilde{\Phi}_{0}\\
\widetilde{\Phi}_{1}\\
\widetilde{\Phi}_{2}\\
\widetilde{\Phi}_{3}%
\end{array}
\right)
\]
and%

\[
\Phi=\mathcal{A}^{-1}[{F}]=\left(
\begin{array}
[c]{rrrr}%
0 & -i & -1 & 0\\
-1 & 0 & 0 & -i\\
1 & 0 & 0 & -i\\
0 & i & -1 & 0
\end{array}
\right)  \left(
\begin{array}
[c]{c}%
\widetilde{F}_{0}\\
\widetilde{F}_{1}\\
\widetilde{F}_{2}\\
\widetilde{F}_{3}%
\end{array}
\right)  .
\]
The following equality is valid
\begin{equation}
D_{\overrightarrow{\alpha}_{sc}}=-\mathcal{A}\gamma_{1}\gamma_{2}\gamma
_{3}\mathcal{D}^{sc}\mathcal{A}^{-1}, \label{sc}%
\end{equation}
where $D_{\overrightarrow{\alpha}_{sc}}=D+M^{\overrightarrow{\alpha}_{sc}}$
and $\overrightarrow{\alpha}_{sc}=-(i\omega e_{1}+(m+\widetilde{\varphi}%
_{sc})e_{2})$. Thus the function $\Phi$ is a solution of the Dirac equation
with scalar potential%
\[
\mathcal{D}^{sc}\Phi=0\qquad\text{in}\quad\Omega
\]
if and only if the function $F=\mathcal{A}\Phi$ is a solution of the equation
\[
D_{\overrightarrow{\alpha}_{sc}}F=0\qquad\text{in}\quad\widetilde{\Omega}.
\]

\subsection{The Dirac equation with electric potential}

The Dirac equation with electric potential has the form%
\[
\mathcal{D}^{el}=\mathcal{D}+i\varphi_{el}\gamma_{0},
\]
where $\varphi_{el}$ is a real-valued function.

We have an equality similar to (\ref{sc}):%
\begin{equation}
D_{\overrightarrow{\alpha}_{el}}=-\mathcal{A}\gamma_{1}\gamma_{2}\gamma
_{3}\mathcal{D}^{el}\mathcal{A}^{-1}, \label{el}%
\end{equation}
where $D_{\overrightarrow{\alpha}_{el}}=D+M^{\overrightarrow{\alpha}_{el}}$
and $\overrightarrow{\alpha}_{el}=-(i(\omega+\varphi_{el})e_{1}+me_{2})$. Thus
the equation $\mathcal{D}^{el}\Phi=0$ is equivalent to the equation
$D_{\overrightarrow{\alpha}_{el}}F=0$.

\subsection{The Dirac equation with pseudoscalar potential}

The Dirac equation with pseudoscalar potential has the form (see, e.g.,
\cite{Thaller})
\[
\mathcal{D}^{ps}=\mathcal{D}+\varphi_{ps}\gamma_{0}\gamma_{5},
\]
where $\varphi_{ps}$ is a real-valued function. We have the following equality
(see \cite{AQA}), similar to (\ref{sc}) and (\ref{el}):%
\[
D+\nu I+M^{\overrightarrow{\beta}}=-\mathcal{A}\gamma_{1}\gamma_{2}\gamma
_{3}\mathcal{D}^{ps}\mathcal{A}^{-1},
\]
where $\nu=-i\widetilde{\varphi}_{ps}$ and $\overrightarrow{\beta}=-(i\omega
e_{1}+me_{2})$.

Suppose that $\overrightarrow{\beta}\notin\mathfrak{S}$, that is $m^{2}%
\neq\omega^{2}$ and denote $S^{\pm}=\frac{1}{2\lambda}M^{(\lambda
\pm\overrightarrow{\beta})}$, where the complex number $\lambda$ is chosen
such that $\lambda^{2}=\overrightarrow{\beta}^{2}$. We have \cite{KrPhys1}
\begin{equation}
D+\nu I+M^{\overrightarrow{\beta}}=S^{+}(D+(\nu+\lambda)I)+S^{-}%
(D+(\nu-\lambda)I). \label{Ps}%
\end{equation}
The operators of multiplication $S^{+}$ and $S^{-}$ are mutually complementary
projection operators on the set of $\mathbb{H}(\mathbb{C})$-valued functions
and commute with the operators in parentheses in (\ref{Ps}). Thus $f$
satisfies the equation
\begin{equation}
(D+\nu I+M^{\overrightarrow{\beta}})f=0 \label{Pseq}%
\end{equation}
if and only if the functions $f^{+}=S^{+}f$ and $f^{-}=S^{-}f$ are solutions
of the equations%
\begin{equation}
(D+\nu+\lambda)f^{+}=0 \label{f+}%
\end{equation}
and
\begin{equation}
(D+\nu-\lambda)f^{-}=0 \label{f-}%
\end{equation}
respectively. In other words, given $f^{+}$ and $f^{-}$ solutions of
(\ref{f+}) and (\ref{f-}), the function $f=S^{+}f^{+}+S^{-}f^{-}$ will be a
solution of (\ref{Pseq}).

Equations (\ref{f+}) and (\ref{f-}) have a quite convenient form for studying,
nevertheless in order to reduce them to the universal form (\ref{1}) we make
one additional step. We have%
\begin{equation}
D+(\nu+\lambda)I=P_{1}^{+}(D+M^{(\nu+\lambda)ie_{1}})+P_{1}^{-}(D-M^{(\nu
+\lambda)ie_{1}}) \label{Ps+}%
\end{equation}
and
\begin{equation}
D+(\nu-\lambda)I=P_{1}^{+}(D+M^{(\nu-\lambda)ie_{1}})+P_{1}^{-}(D-M^{(\nu
-\lambda)ie_{1}}). \label{Ps-}%
\end{equation}
Thus $f^{+}$ and $f^{-}$ are solutions of (\ref{f+}) and (\ref{f-})
respectively if and only if the functions $f^{++}=P_{1}^{+}f^{+}$ and
$f^{-+}=P_{1}^{-}f^{+}$ are solutions of the equations%
\begin{equation}
(D+M^{(\nu+\lambda)ie_{1}})f^{++}=0, \label{++}%
\end{equation}%
\begin{equation}
(D-M^{(\nu+\lambda)ie_{1}})f^{-+}=0, \label{-+}%
\end{equation}
and the functions $f^{+-}=P_{1}^{+}f^{-}$ and $f^{--}=P_{1}^{-}f^{-}$ are
solutions of the equations%
\begin{equation}
(D+M^{(\nu-\lambda)ie_{1}})f^{+-}=0, \label{+-}%
\end{equation}%
\[
(D-M^{(\nu-\lambda)ie_{1}})f^{--}=0.
\]
The obtained result we resume in the following statement.

\begin{proposition}
For $\overrightarrow{\beta}\notin\mathfrak{S}$ and $\lambda^{2}%
=\overrightarrow{\beta}^{2}$ the following equality is valid%
\begin{align*}
D+\nu I+M^{\overrightarrow{\beta}}  &  =P_{1}^{+}S^{+}(D+M^{(\nu
+\lambda)ie_{1}})+P_{1}^{-}S^{+}(D-M^{(\nu+\lambda)ie_{1}})\\
&  +P_{1}^{+}S^{-}(D+M^{(\nu-\lambda)ie_{1}})+P_{1}^{-}S^{-}(D-M^{(\nu
-\lambda)ie_{1}})
\end{align*}
which implies
\begin{align*}
\ker(D+\nu I+M^{\overrightarrow{\beta}})  &  =P_{1}^{+}S^{+}\ker
(D+M^{(\nu+\lambda)ie_{1}})\oplus P_{1}^{-}S^{+}\ker(D-M^{(\nu+\lambda)ie_{1}%
})\\
&  \oplus P_{1}^{+}S^{-}\ker(D+M^{(\nu-\lambda)ie_{1}})\oplus P_{1}^{-}%
S^{-}\ker(D-M^{(\nu-\lambda)ie_{1}}),
\end{align*}
where $\ker$ means the set of null-solutions in a domain of interest
$\Omega\subset\mathbb{R}^{3}$.
\end{proposition}

In this way equation (\ref{Pseq}) and hence the Dirac equation with
pseudoscalar potential reduce to four equations of the form (\ref{1}).

\subsection{Force-free magnetic fields}

Force-free magnetic fields appear as an important class of special solutions
of nonlinear equations of magnetohydrodynamics and are intensively studied in
different branches of modern physics (see, e.g., \cite{Stratis}, \cite{Clegg},
\cite{Feng}, \cite{Kaiser}, \cite{Kamien}, \cite{Lifschitz}, \cite{Priest},
\cite{Schnack}, \cite{Yoshida}, \cite{Zag}). They are characterized by the
following pair of equations%
\begin{equation}
\operatorname{div}\mathbf{B}=0 \label{ff1}%
\end{equation}
and%
\begin{equation}
\operatorname{rot}\mathbf{B}+\nu\mathbf{B}=0, \label{ff2}%
\end{equation}
where $\nu$ is a scalar function. This system can be rewritten in the form%
\begin{equation}
(D+\nu)\mathbf{B}=0. \label{ff}%
\end{equation}
It will be more convenient for us to extend the class of solutions of
(\ref{ff}) and to consider not only its purely vectorial solutions but more
general complete biquaternionic functions. Thus we consider the equation%
\begin{equation}
(D+\nu)f=0, \label{ffq}%
\end{equation}
where $f$ is an $\mathbb{H}(\mathbb{C})$-valued function. Solutions of
(\ref{ff1}), (\ref{ff2}) represent a subset of solutions of (\ref{ffq}) which
fulfil the additional requirement $\operatorname{Sc}f=0$.

Now by analogy with the preceding subsection we obtain that $f$ is a solution
of (\ref{ffq}) if and only if the functions $f^{+}=P_{1}^{+}f$ and
$f^{-}=P_{1}^{-}f$ are solutions of the equations $(D+M^{i\nu e_{1}})f^{+}=0$
\ and $(D-M^{i\nu e_{1}})f^{-}=0$ respectively.

Thus the system (\ref{ff1}), (\ref{ff2}) reduces to a pair of equations of the
form (\ref{1}).

\subsection{Maxwell's equations for slowly changing media}

Consider first the general Maxwell system%
\begin{equation}
\operatorname{rot}\mathbf{H}=\varepsilon\partial_{t}\mathbf{E}+\mathbf{j,}
\label{Min1}%
\end{equation}%
\begin{equation}
\operatorname{rot}\mathbf{E}=-\mu\partial_{t}\mathbf{H}, \label{Min2}%
\end{equation}%
\begin{equation}
\operatorname{div}(\varepsilon\mathbf{E)}=\rho, \label{Min3}%
\end{equation}%
\begin{equation}
\operatorname{div}\mathbf{(}\mu\mathbf{H)}=0, \label{Min4}%
\end{equation}
where $\varepsilon$ and $\mu$ are assumed to be functions of spatial
coordinates only. It can be rewritten in the following form \cite{KrZAA02},
\cite{AQA}%
\begin{equation}
(D+M^{\overrightarrow{\varepsilon}})\overrightarrow{E}=-\frac{1}{c}%
\partial_{t}\overrightarrow{H}-\frac{\rho}{\sqrt{\varepsilon}}, \label{Minq1}%
\end{equation}
and%
\begin{equation}
(D+M^{\overrightarrow{\mu}})\overrightarrow{H}=\frac{1}{c}\partial
_{t}\overrightarrow{E}+\sqrt{\mu}\mathbf{j}, \label{Minq2}%
\end{equation}
where $\overrightarrow{E}=\sqrt{\varepsilon}\mathbf{E,}$ $\overrightarrow
{H}=\sqrt{\mu}\mathbf{H}$, $c=1/\sqrt{\varepsilon\mu}$, $\overrightarrow
{\varepsilon}=\frac{\operatorname{grad}\sqrt{\varepsilon}}{\sqrt{\varepsilon}%
}$ and $\overrightarrow{\mu}=\frac{\operatorname{grad}\sqrt{\mu}}{\sqrt{\mu}}$.

In a sourceless time-harmonic situation we obtain the equations
\begin{equation}
D_{\overrightarrow{\varepsilon}}\overrightarrow{E}=i\nu\overrightarrow
{H}\qquad\text{and\qquad}D_{\overrightarrow{\mu}}\overrightarrow{H}%
=-i\nu\overrightarrow{E}. \label{Mm}%
\end{equation}
Here $\nu=\omega/c$.

The medium is said to be slowly changing when its properties change
appreciably over distances much greater than the wavelength \cite{Babich,
Vinogradova}. Usually this is associated with the possibility of reducing the
Maxwell equations (\ref{Mm}) to the Helmholtz equations%
\[
(\Delta+\nu^{2})\overrightarrow{E}=0\qquad\text{and\qquad}(\Delta+\nu
^{2})\overrightarrow{H}=0.
\]
It is easy to check that such a reduction is possible if only $\left|
\overrightarrow{\varepsilon}\right|  $ and $\left|  \overrightarrow{\mu
}\right|  $ are considered as relatively very small and the terms containing
the vectors $\overrightarrow{\varepsilon}$ and $\overrightarrow{\mu}$ are
supposed to be negligible. Then (\ref{Mm}) take the form%
\[
D\overrightarrow{E}=i\nu\overrightarrow{H}\qquad\text{and\qquad}%
D\overrightarrow{H}=-i\nu\overrightarrow{E}%
\]
and can be diagonalized. For the functions $\overrightarrow{\varphi
}=\overrightarrow{E}+i\overrightarrow{H}$ and $\overrightarrow{\psi
}=\overrightarrow{E}-i\overrightarrow{H}$ we obtain the equations%
\[
(D-\nu)\overrightarrow{\varphi}=0\qquad\text{and\qquad}(D+\nu)\overrightarrow
{\psi}=0.
\]
Now, by analogy with equation (\ref{ff}) each of these two equations can be
rewritten in the form (\ref{1}).

\subsection{The static Maxwell system}

When the vectors of the electromagnetic field do not depend on time, from
(\ref{Minq1}) and (\ref{Minq2}) we obtain two independent equations
\begin{equation}
(D+M^{\overrightarrow{\varepsilon}})\overrightarrow{E}=-\frac{\rho}%
{\sqrt{\varepsilon}}, \label{statE}%
\end{equation}
and%
\begin{equation}
(D+M^{\overrightarrow{\mu}})\overrightarrow{H}=\sqrt{\mu}\mathbf{j}.
\label{statH}%
\end{equation}
In a sourceless situation both equations reduce to (\ref{1}).

\subsection{Two important cases}

Let us resume the results presented in this section. We see that the
considered here six physical models reduce to equation (\ref{1}). In the case
of the static Maxwell system $\overrightarrow{\alpha}$ is a gradient of some
scalar function. In all other cases in general $\overrightarrow{\alpha}$ must
not be necessarily a gradient, though such a possibility is not excluded. In
the first five models only one of the components of $\overrightarrow{\alpha}$
is a function while the other two are constants. Thus we are interested
basically in the following two situations.

\begin{enumerate}
\item $\overrightarrow{\alpha}$ is a gradient of some scalar function
$\varphi$: $\overrightarrow{\alpha}=\nabla\varphi$.

\item $\overrightarrow{\alpha}$ has the form $\overrightarrow{\alpha}%
=\alpha_{1}(x_{1},x_{2},x_{3})e_{1}+\alpha_{2}e_{2}+\alpha_{3}e_{3}$, where
$\alpha_{1}$ is a complex valued function and $\alpha_{2}$, $\alpha_{3}$ are
complex constants (of course, when $\alpha_{1}=\alpha_{1}(x_{1})$ we have the
first case again).
\end{enumerate}

\section{Factorization of the Schr\"{o}dinger operator}

\subsection{Scalar Schr\"{o}dinger equations}

Consider the Schr\"{o}dinger operator $-\Delta+v$ applied to a scalar function
$\phi$. Here $v$ is some complex valued function. Let $\overrightarrow{\alpha
}$ be a purely vectorial biquaternion valued function such that%
\begin{equation}
D\overrightarrow{\alpha}+(\overrightarrow{\alpha})^{2}=-v. \label{Ricc}%
\end{equation}
Then as was observed in \cite{Swansolo} and \cite{Swan}, the following
equality is valid%
\begin{equation}
(-\Delta+v)\phi=(D+M^{\overrightarrow{\alpha}})(D-M^{\overrightarrow{\alpha}%
})\phi. \label{scfact}%
\end{equation}
This equality gives a certain relation between solutions of (\ref{1}) and
null-solutions of the Schr\"{o}dinger operator. Moreover, a fundamental
solution of the operator $D_{\overrightarrow{\alpha}}$ can be obtained if a
fundamental solution of the Schr\"{o}dinger operator is given. In some cases,
for instance when $\overrightarrow{\alpha}=\alpha_{1}(x_{1})e_{1}$, this leads
to construction of integral representations for solutions of (\ref{1}) (see
\cite{KrBelt}).

Equation (\ref{Ricc}) represents a generalization of the famous Riccati
differential equation. In \cite{KKW} (see also \cite{AQA}) it was shown that
it is not merely a formal resemblance. On the contrary, the well known Euler
theorems, the Weyr theorem as well as the Picard theorem descibing the unique
properties of the Riccati equation were generalized for equation (\ref{Ricc}).

Equation (\ref{Ricc}) necessarily implies that $\overrightarrow{\alpha}$ is a
gradient, because from the vector part of (\ref{Ricc}) we have that
$\operatorname{rot}\overrightarrow{\alpha}=0$. Thus it can be useful only for
the first situation from Subsection 3.7.

Let us consider the product of the operators $D_{\overrightarrow{\alpha}}$ and
$D_{-\overrightarrow{\alpha}}$ in application to a twice differentiable
biquaternion valued function $u$ and for any differentiable purely vectorial
biquaternion $\overrightarrow{\alpha}$. We have%
\begin{align*}
D_{\overrightarrow{\alpha}}D_{-\overrightarrow{\alpha}}u  &
=(D+M^{\overrightarrow{\alpha}})(D-M^{\overrightarrow{\alpha}})\sum_{k=0}%
^{3}u_{k}e_{k}\\
&  =\sum_{k=0}^{3}(M^{e_{k}}(D+M^{\overrightarrow{\alpha}^{(k)}}%
)(D-M^{\overrightarrow{\alpha}^{(k)}})u_{k}).
\end{align*}
For each component $u_{k}$ we use (\ref{scfact}) and note that
$(\overrightarrow{\alpha}^{(k)})^{2}=\overrightarrow{\alpha}^{2}$. We obtain%
\begin{equation}
D_{\overrightarrow{\alpha}}D_{-\overrightarrow{\alpha}}u=\sum_{k=0}%
^{3}(-\Delta u_{k}-\overrightarrow{\alpha}^{2}u_{k}-(D\overrightarrow{\alpha
}^{(k)})u_{k})e_{k}. \label{fact}%
\end{equation}
From (\ref{fact}) we see that the equation $D_{\overrightarrow{\alpha}%
}D_{-\overrightarrow{\alpha}}u=0$ is equivalent to four scalar Schr\"{o}dinger
equations if only $D\overrightarrow{\alpha}^{(k)}$ is scalar for any
$k=0,1,2,3$. It is easy to verify that it is possible if only $\overrightarrow
{\alpha}$ has the following form%
\begin{equation}
\overrightarrow{\alpha}=\alpha_{1}(x_{1})e_{1}+\alpha_{2}(x_{2})e_{2}%
+\alpha_{3}(x_{3})e_{3}. \label{alpha}%
\end{equation}
Let us consider this case in detail. First of all we notice that for the Dirac
operator with scalar, electric or pseudoscalar potential this restriction on
$\overrightarrow{\alpha}$ implies that the potential is an arbitrary function
of one spatial coordinate. In the case of force-free magnetic fields
(Subsection 3.4) the proportionality factor $\nu$ is a function of one
variable. In the case of the electromagnetic field in a slowly changing medium
(Subsection 3.5) the wave number $\nu$ is a function of one variable. Such
media are known as stratified media.

For the static Maxwell system $\operatorname{div}(\varepsilon\mathbf{E})=0$,
and $\operatorname{rot}\mathbf{E}=0$ we obtain that it is equivalent to
(\ref{1}) with $\overrightarrow{\alpha}$ having the form (\ref{alpha}) iff the
permittivity $\varepsilon$ has the following form $\varepsilon=\varepsilon
_{1}(x_{1})\varepsilon_{2}(x_{2})\varepsilon_{3}(x_{3})$, where $\varepsilon
_{1}$, $\varepsilon_{2}$ and $\varepsilon_{3}$ are arbitrary functions.
$\alpha_{k}$ are related to $\varepsilon_{k}$ in the following way $\alpha
_{k}=(\partial_{k}\varepsilon_{k})/(2\varepsilon_{k})$, $k=1,2,3$.

Thus for all considered physical models $\overrightarrow{\alpha}$ in the form
(\ref{alpha}) corresponds to quite interesting, nontrivial situations. Let us
study in detail the structure of solutions of equation (\ref{1}) when
$\overrightarrow{\alpha}$ has the form (\ref{alpha}).

Denote
\begin{equation}
v_{k}=-D\overrightarrow{\alpha}^{(k)}-\overrightarrow{\alpha}^{2}, \label{vk}%
\end{equation}
and%
\begin{equation}
w_{k}=D\overrightarrow{\alpha}^{(k)}-\overrightarrow{\alpha}^{2},\qquad
k=0,1,2,3. \label{wk}%
\end{equation}

\begin{proposition}
Let $f$ be a solution of (\ref{1}) with $\overrightarrow{\alpha}$ having the
form (\ref{alpha}). Then the components $f_{k}$ are solutions of the
Schr\"{o}dinger equations
\begin{equation}
(-\Delta+w_{k})f_{k}=0,\qquad k=0,1,2,3. \label{Schrw}%
\end{equation}
\end{proposition}

\begin{proof}
Assume that $f$ is a solution of (\ref{1}). Then considering the equation
$D_{-\overrightarrow{\alpha}}D_{\overrightarrow{\alpha}}f=0$ we arrive at the
equations (\ref{Schrw}) for the components $f_{k}$.
\end{proof}

The following statement gives us a method for constructing exact solutions of
(\ref{1}) having obtained solutions of the corresponding Schr\"{o}dinger equations.

\begin{proposition}
Let $\overrightarrow{\alpha}$ be of the form (\ref{alpha}) and four scalar
functions $g_{k}$, $k=0,1,2,3$ satisfy the following equations
\begin{equation}
(-\Delta+v_{k})g_{k}=0. \label{Schrv}%
\end{equation}
Then the function
\begin{equation}
f=(D-M^{\overrightarrow{\alpha}})g \label{solut}%
\end{equation}
is a solution of (\ref{1}), where $g=\sum_{k=0}^{3}g_{k}e_{k}$.
\end{proposition}

\begin{proof}
This is an immediate consequence of (\ref{fact}).
\end{proof}

\begin{example}
\label{Ex1}Let $\overrightarrow{\alpha}=\sum_{k=1}^{3}\frac{1}{x_{k}-b_{k}%
}e_{k}$, where $b_{k}$ are arbitrary complex constants. We have
\begin{equation}
v_{0}=-D\overrightarrow{\alpha}-\overrightarrow{\alpha}^{2}=0. \label{pot0}%
\end{equation}
This means that taking any scalar harmonic function $g_{0}$ we are able to
construct a solution of (\ref{1}) in the form $f=(D-M^{\overrightarrow{\alpha
}})g_{0}$. Calculating $v_{1}$, $v_{2}$ and $v_{3}$ according to (\ref{vk}) we
obtain%
\begin{equation}
v_{1}=2\left(  \frac{1}{(x_{2}-b_{2})^{2}}+\frac{1}{(x_{3}-b_{3})^{2}}\right)
, \label{pot1}%
\end{equation}%
\begin{equation}
v_{2}=2\left(  \frac{1}{(x_{1}-b_{1})^{2}}+\frac{1}{(x_{3}-b_{3})^{2}}\right)
, \label{pot2}%
\end{equation}%
\begin{equation}
v_{3}=2\left(  \frac{1}{(x_{1}-b_{1})^{2}}+\frac{1}{(x_{2}-b_{2})^{2}}\right)
. \label{pot3}%
\end{equation}
\end{example}

We will not try to find general solutions of the Schr\"{o}dinger equations
with these potentials. We show instead how one can always obtain a class of
exact solutions of (\ref{1}) and of the corresponding Schr\"{o}dinger
equations (\ref{Schrv}) when $\overrightarrow{\alpha}$ has the form
(\ref{alpha}).

Let us look for a one-component solution of (\ref{1}): $f=f_{k}e_{k}$. We have
$Df_{k}e_{k}+f_{k}\overrightarrow{\alpha}^{(k)}e_{k}=0$. That is
\begin{equation}
\frac{\nabla f_{k}}{f_{k}}=-\overrightarrow{\alpha}^{(k)}. \label{alphak}%
\end{equation}
It is easy to see that for $\overrightarrow{\alpha}$ of the form (\ref{alpha})
the functions $f_{k}$ are
\[
f_{0}=e^{-(\Lambda_{1}+\Lambda_{2}+\Lambda_{3})},\quad f_{1}=e^{-\Lambda
_{1}+\Lambda_{2}+\Lambda_{3}},\quad f_{2}=e^{\Lambda_{1}-\Lambda_{2}%
+\Lambda_{3}},\quad f_{3}=e^{\Lambda_{1}+\Lambda_{2}-\Lambda_{3}},
\]
where $\Lambda_{k}=\Lambda_{k}(x_{k})$, $k=1,2,3$ is an antiderivative of
$\alpha_{k}$. Thus the function $\sum_{k=0}^{3}c_{k}f_{k}e_{k}$ is a solution
of (\ref{1}), where $c_{k}$ are arbitrary complex constants.

Concerning the Schr\"{o}dinger equations (\ref{Schrv}) it is useful to
consider the quaternionic Riccati equation (\ref{Ricc}). Any solution of it
can be represented in the form $\overrightarrow{\alpha}=\frac{\nabla\varphi
}{\varphi}$, where $\varphi$ is a solution of the equation $-\Delta
\varphi+v\varphi=0$ (see \cite{KKW}, \cite{AQA}). For $\overrightarrow{\alpha
}^{(k)}$ from (\ref{alphak}) we obtain that $\overrightarrow{\alpha}%
^{(k)}=\frac{\nabla\varphi_{k}}{\varphi_{k}}$, where $\varphi_{k}=1/f_{k}$ and
consequently the functions $\varphi_{k}$ are solutions of (\ref{Schrv}) for a
corresponding $k$.

Let us see what are the functions $f_{k}$ and $\varphi_{k}$ in Example
\ref{Ex1}.

\begin{example}
For $\overrightarrow{\alpha}$ from the preceding example we have
\[
f_{0}=\frac{1}{(x_{1}-b_{1})(x_{2}-b_{2})(x_{3}-b_{3})},\quad f_{1}%
=\frac{(x_{2}-b_{2})(x_{3}-b_{3})}{x_{1}-b_{1}},
\]%
\[
f_{2}=\frac{(x_{1}-b_{1})(x_{3}-b_{3})}{x_{2}-b_{2}},\quad f_{3}%
=\frac{(x_{1}-b_{1})(x_{2}-b_{2})}{x_{3}-b_{3}}.
\]
Hence the function $f=\sum_{k=0}^{3}c_{k}f_{k}e_{k}$ is a solution of
(\ref{1}) and the functions
\[
\varphi_{0}=(x_{1}-b_{1})(x_{2}-b_{2})(x_{3}-b_{3}),\quad\varphi
_{1}=\frac{x_{1}-b_{1}}{(x_{2}-b_{2})(x_{3}-b_{3})},
\]%
\[
\varphi_{2}=\frac{x_{2}-b_{2}}{(x_{1}-b_{1})(x_{3}-b_{3})},\quad\varphi
_{3}=\frac{x_{3}-b_{3}}{(x_{1}-b_{1})(x_{2}-b_{2})}%
\]
are solutions of (\ref{Schrv}) with potentials (\ref{pot0})-(\ref{pot3}) respectively.
\end{example}

\begin{proposition}
\label{gsol}Let $\overrightarrow{\alpha}$ be of the form (\ref{alpha}) and
$f=(D-M^{\overrightarrow{\alpha}})g$ a solution of (\ref{1}), where
$g=\sum_{k=0}^{3}g_{k}e_{k}$. Then necessarily $g_{k}$ are solutions of
(\ref{Schrv}) respectively.
\end{proposition}

\begin{proof}
Consider
\begin{align*}
(D+M^{\overrightarrow{\alpha}})f  &  =(D+M^{\overrightarrow{\alpha}%
})(D-M^{\overrightarrow{\alpha}})\sum_{k=0}^{3}g_{k}e_{k}\\
&  =\sum_{k=0}^{3}(M^{e_{k}}(D+M^{\overrightarrow{\alpha}^{(k)}}%
)(D-M^{\overrightarrow{\alpha}^{(k)}})g_{k})\\
&  =\sum_{k=0}^{3}(M^{e_{k}}(-\Delta+v_{k})g_{k}).
\end{align*}
From assumption\ of the proposition we obtain that $(-\Delta+v_{k})g_{k}$,
$k=0,1,2,3$.
\end{proof}

Let $\Omega$ be a domain in $\mathbb{R}^{3}$ which in particular may coincide
with the whole space. Let $F(\Omega)$ and $G(\Omega)$ be some functional
spaces. The $\mathbb{H}(\mathbb{C})$-valued function $f$ is said to belong to
a functional space if each of its components $f_{k}$ belongs to it.

\begin{proposition}
\label{exist}Let $\overrightarrow{\alpha}$ be of the form (\ref{alpha}).
Assume that the equation $(-\Delta+w_{k}(\mathbf{x}))u(\mathbf{x}%
)=\mu(\mathbf{x})$, $\mathbf{x}\in\Omega$, $k=0,1,2,3$ is solvable for any
$\mu\in F(\Omega)$ and the solution $u$ belongs to $G(\Omega)$. Then the
equation%
\begin{equation}
(D-M^{\overrightarrow{\alpha}})g=f \label{inhomog}%
\end{equation}
is solvable for any $f\in F(\Omega)$ and the solution $g$ belongs to
$\operatorname{im}D_{\overrightarrow{\alpha}}(G(\Omega))$.
\end{proposition}

\begin{proof}
Let $u_{k}$, $k=0,1,2,3$ be solutions of the equations $(-\Delta+w_{k}%
)u_{k}=f_{k}$. Then $g=(D+M^{\overrightarrow{\alpha}})\sum_{k=0}^{3}u_{k}%
e_{k}$ is a solution of (\ref{inhomog}).
\end{proof}

\begin{remark}
We do not specify the functional spaces here because the results on
solvability of the inhomogeneous Schr\"{o}dinger equation are numerous and
correspond to very different situations. Let us give an example.
\end{remark}

\begin{example}
Let $\Omega=\mathbb{R}^{3}$ and $w$ has the form
\begin{equation}
w=\widehat{w}-m^{2}, \label{potw}%
\end{equation}
where $\widehat{w}\in C_{0}^{\infty}(\mathbb{R}^{3})$, $m>0$. The equation
$(-\Delta+w)u=\mu$ \ is uniquely solvable\ \cite{Egorov} for any $\mu\in
L_{2,a}$ and $u\in H_{loc}^{2}(\mathbb{R}^{3})$, $u=O(\frac{1}{\left|
\mathbf{x}\right|  })$, $\frac{\partial u}{\partial\left|  \mathbf{x}\right|
}-imu=o(\frac{1}{\left|  \mathbf{x}\right|  })$ for $\left|  \mathbf{x}%
\right|  \rightarrow\infty$. Here $L_{2,a}$ denotes the space of square
integrable functions with support in a ball of radius $a$.

Thus from Proposition \ref{exist} we have the solvability of (\ref{inhomog})
for any $\overrightarrow{\alpha}$ such that $w_{k}$ defined by (\ref{wk}) have
the form (\ref{potw}), for example, for $\overrightarrow{\alpha}=\sum
_{k=1}^{3}\beta_{k}(x_{k})+im_{k}$, where $\beta_{k}\in C_{0}^{\infty
}(\mathbb{R}^{3})$ and $m_{k}>0$.
\end{example}

\begin{proposition}
Under the conditions of Proposition \ref{exist} any solution of (\ref{1}) from
$F(\Omega)$ has the form
\begin{equation}
f=(D-M^{\overrightarrow{\alpha}})g, \label{formf}%
\end{equation}
where $g=\sum_{k=0}^{3}g_{k}e_{k}$.and $g_{k}$ satisfy the equations
(\ref{Schrv}) in $\Omega$.
\end{proposition}

\begin{proof}
From Proposition \ref{exist} it follows that $f$ can be represented in the
form (\ref{formf}), where $g\in\operatorname{im}D_{\overrightarrow{\alpha}%
}(G(\Omega))$. From Proposition \ref{gsol} we obtain that $g_{k}$ are
solutions of (\ref{Schrv}).
\end{proof}

For the Schr\"{o}dinger operator there are developed different approaches for
obtaining asymptotic fundamental solutions under some additional assumptions.
A fundamental solution can be used for construction of a right-inverse
operator for the Schr\"{o}dinger operator which gives a possibility to solve
inhomogeneous equations. In the following proposition we show that having
constructed \ right-inverse operators for the Schr\"{o}dinger operators
$-\Delta+v_{k}$, $k=0,1,2,3$, one can construct a right-inverse operator for
$D_{\overrightarrow{\alpha}}$.

\begin{proposition}
Let $\overrightarrow{\alpha}$ be of the form (\ref{alpha}) and $T_{k}$ such
operators that for any $\varphi\in F(\Omega)$: $(-\Delta+v_{k})T_{k}%
\varphi=\varphi$ in $\Omega$, $k=0,1,2,3$. Then for any $f=\sum_{k=0}^{3}%
f_{k}e_{k}\in F(\Omega)$ we have $D_{\overrightarrow{\alpha}}%
T_{\overrightarrow{\alpha}}f=f$ in $\Omega,$ where $T_{\overrightarrow{\alpha
}}f=(D-M^{\overrightarrow{\alpha}})\left(  \sum_{k=0}^{3}(T_{k}f_{k}%
)e_{k}\right)  $.
\end{proposition}

\begin{proof}
Consider%
\begin{align*}
(D+M^{\overrightarrow{\alpha}})T_{\overrightarrow{\alpha}}f  &  =\sum
_{k=0}^{3}(M^{e_{k}}(D+M^{\overrightarrow{\alpha}^{(k)}})(D-M^{\overrightarrow
{\alpha}^{(k)}})(T_{k}f_{k}))\\
&  =\sum_{k=0}^{3}(M^{e_{k}}(-\Delta+v_{k})T_{k}f_{k})=\sum_{k=0}^{3}%
f_{k}e_{k}=f.
\end{align*}
\end{proof}

\subsection{Schr\"{o}dinger equations with quaternionic potentials}

Let us assume that $\overrightarrow{\alpha}$ has the form
\begin{equation}
\overrightarrow{\alpha}=\alpha_{1}(x_{1},x_{2},x_{3})e_{1}+\alpha_{2}%
e_{2}+\alpha_{3}e_{3} \label{alpha2}%
\end{equation}
where $\alpha_{1}$ is an arbitrary complex valued differentiable function and
$\alpha_{2}$, $\alpha_{3}$ are complex constants. In Section 3 we saw that the
first five physical models considered here correspond to this situation.

Note that in this case $D\overrightarrow{\alpha}=D\overrightarrow{\alpha
}^{(1)}=-D\overrightarrow{\alpha}^{(2)}=-D\overrightarrow{\alpha}^{(3)}$.
Hence from (\ref{fact}) we obtain
\[
D_{\overrightarrow{\alpha}}D_{-\overrightarrow{\alpha}}u=-\Delta
u-\overrightarrow{\alpha}^{2}u-(D\overrightarrow{\alpha})(u_{0}e_{0}%
+u_{1}e_{1}-u_{2}e_{2}-u_{3}e_{3}),
\]
where $u=\sum_{k=0}^{3}u_{k}e_{k}$.

Denote $Cu=-e_{1}ue_{1}=u_{0}e_{0}+u_{1}e_{1}-u_{2}e_{2}-u_{3}e_{3}$,
$Au=-\Delta u-\overrightarrow{\alpha}^{2}u$ and $Bu=-(D\overrightarrow{\alpha
})u$. That is
\begin{equation}
D_{\overrightarrow{\alpha}}D_{-\overrightarrow{\alpha}}=A+BC \label{factq1}%
\end{equation}
and
\[
D_{-\overrightarrow{\alpha}}D_{\overrightarrow{\alpha}}=A-BC.
\]
Denote $Q^{\pm}=\frac{1}{2}(I\pm ie_{1}C)$.

\begin{proposition}
Solutions of equation (\ref{1}) (with $\overrightarrow{\alpha}$ of the form
(\ref{alpha2})) have the form $f=D_{-\overrightarrow{\alpha}}u$, where
$u=Q^{+}v+Q^{-}w$ and $v$, $w$ are solutions of the following Schr\"{o}dinger
equations with quaternionic potentials%
\begin{equation}
(A+Bie_{1})v=0 \label{ABC+}%
\end{equation}
and
\begin{equation}
(A-Bie_{1})w=0 \label{ABC-}%
\end{equation}
respectively.
\end{proposition}

\begin{proof}
First, we notice that $Q^{\pm}B=BQ^{\pm}$ and $Q^{\pm}C=\frac{1}{2}(C\pm
ie_{1})=\frac{1}{2}(Cie_{1}\pm I)ie_{1}=\pm Q^{\pm}ie_{1}$.

Applying $Q^{+}$ and $Q^{-}$ to the equation%
\begin{equation}
(A+BC)u=0 \label{ABC}%
\end{equation}
we see that it is equivalent to the pair of equations $(A+Bie_{1})Q^{+}u=0$,
$(A-Bie_{1})Q^{-}u=0,$ and $Q^{+}$, $Q^{-}$ commute with the operators in
parentheses and represent a pair of mutually complementary projection
operators on the space of $\mathbb{H}(\mathbb{C})$-valued functions. Thus we
have that $u$ is a solution of (\ref{ABC}) iff $u=Q^{+}v+Q^{-}w$ and $v$, $w$
are solutions of (\ref{ABC+}) and (\ref{ABC-}) respectively. Now using
(\ref{factq1}) we finish the proof.
\end{proof}

Thus solution of (\ref{1}) with $\overrightarrow{\alpha}$ of the form
(\ref{alpha2}) reduces to solution of the Schr\"{o}dinger equations with
quaternionic potentials ((\ref{ABC+}) and (\ref{ABC-}) which can be rewritten
in a more explicit form as follows%
\begin{equation}
(-\Delta-(\overrightarrow{\alpha}^{2}-iD\alpha_{1})I)v=0 \label{ABC+1}%
\end{equation}
and
\begin{equation}
(-\Delta-(\overrightarrow{\alpha}^{2}+iD\alpha_{1})I)w=0. \label{ABC-1}%
\end{equation}

\begin{remark}
In the case when $\alpha_{1}$ does not depend on $x_{1}$: $\alpha_{1}%
=\alpha_{1}(x_{2},x_{3})$ it is easy to see that equations (\ref{ABC+1}) and
(\ref{ABC-1}) are not independent. We have that if $v$ is a solution of
(\ref{ABC+1}) then $w=ie_{1}v$ is a solution of (\ref{ABC-1}) and vice versa.
Using this fact we obtain a one-to-one correspondence between solutions of
(\ref{ABC}) and solutions of (\ref{ABC+1}). Such a correspondence is given by
the operator%
\[
\Pi=\frac{1}{2}(I+ie_{1}-C+ie_{1}C).
\]
It can be verified immediately that $\Pi^{2}=I$ and that $u$ is a solution of
(\ref{ABC}) if and only if $v=\Pi u$ is a solution of (\ref{ABC+1}) and vice
versa. That is when $\overrightarrow{\alpha}=\alpha_{1}(x_{2},x_{3}%
)e_{1}+\alpha_{2}e_{2}+\alpha_{3}e_{3}$ where $\alpha_{2}$, $\alpha_{3}$ are
constants, solution of (\ref{1}) reduces to solution of one Schr\"{o}dinger
equation with quaternionic potential (\ref{ABC+1}).
\end{remark}

\begin{remark}
Some classes of solutions of (\ref{ABC+1}) and (\ref{ABC-1}) can be obtained
reducing the equations to scalar Schr\"{o}dinger's equations using the
following idea (proposed in \cite{KrBirk} in another setting). Consider
equation (\ref{ABC+1}). If $(iD\alpha_{1}-\overrightarrow{\alpha}^{2}%
)\in\mathfrak{S}$ or $D\alpha_{1}\in\mathfrak{S}$ then we can look for
solutions of (\ref{ABC+1}) in the form $v=(iD\alpha_{1}+\overrightarrow
{\alpha}^{2})f$ or $v=(D\alpha_{1})f$ respectively where $f$ is an unknown
function. In the first case the equation reduces to the Laplace equation
$\Delta v=0$ and in the second to the Schr\"{o}dinger equation $(\Delta
+\overrightarrow{\alpha}^{2})v=0$.

Suppose that neither of these is the case. Then denote $\overrightarrow{\beta
}=iD\alpha_{1}$ and introduce $\beta_{0}$ as a scalar square root of
$\overrightarrow{\beta}^{2}$. The complex quaternions $\beta=\beta
_{0}+\overrightarrow{\beta}$ and $\overline{\beta}=\beta_{0}-\overrightarrow
{\beta}$ are conjugate zero divisors. Equation (\ref{ABC+1}) then can be
rewritten as follows%
\[
(-\Delta-(\beta_{0}+\overrightarrow{\alpha}^{2})+\beta)v=0.
\]
Looking for its solutions of the form $v=\overline{\beta}f$ we reduce it to
the scalar Schr\"{o}dinger equation $(\Delta+(\beta_{0}+\overrightarrow
{\alpha}^{2}))v=0$.
\end{remark}

\section{Conclusions}

We have shown that an ample class of physically meaningful partial
differential systems of first order are equivalent to a single quaternionic
equation which in its turn reduces in general to a Schr\"{o}dinger equation
with quaternionic potential, and in some situations considered in Subsection
4.1 this last can be diagonalized. The rich variety of methods developed for
different problems corresponding to the Schr\"{o}dinger equation can be
applied to the systems considered in the present work.

\end{document}